\documentclass[prd,article]{revtex4}

\usepackage{mathrsfs}
\usepackage{amsmath,amssymb,amsbsy,amsfonts}
\usepackage{graphicx}
\usepackage{amsmath,amssymb,amsbsy,amsfonts}
\usepackage[usenames]{color}
\usepackage{natbib}

\def\n{{\rm n}}
\def\x{{\rm x}}
\def\y{{\rm y}}

\newcommand{\ncd}{\newcommand}
\ncd{\mrm} {\mathrm}
\ncd{\beq} {\begin{equation}}
\ncd{\eeq} {\end{equation}}
\ncd{\bea} {\begin{eqnarray}}
\ncd{\eea} {\end{eqnarray}}

\def\eg{{\it eg.~}}  \def\ie{{\it i.e.~}}

\ncd{\nx}{n_\x^2}
\ncd{\ny}{n_\y^2}
\ncd{\nxy}{n_{\x\y}^2}
\ncd{\nxbar}{\bar{n}_\x^2}
\ncd{\nybar}{\bar{n}_\y^2}
\ncd{\nxybar}{\bar{n}_{\xY}^2}
\ncd{\nyxbar}{\bar{n}_{\y\x}^2}
\ncd{\knx}{(k\cdot n_\x)}
\ncd{\kny}{(k\cdot n_\y)}
\ncd{\knxbar}{(k\cdot \bar{n}_\x)}
\ncd{\knybar}{(k\cdot \bar{n}_\y)}
\ncd{\Bn}{\mathcal{B}^\n}
\ncd{\BX}{\mathcal{B}^\x}
\ncd{\BY}{\mathcal{B}^\y}
\ncd{\AXY}{\mathcal{A}^{\x\y}}
\ncd{\AXYXY}{\frac{\partial\AXY}{\partial\nxy}}
\ncd{\AXYX}{\frac{\partial\AXY}{\partial\nx}}
\ncd{\AXYY}{\frac{\partial\AXY}{\partial\ny}}
\ncd{\BXX}{\frac{\partial\BX}{\partial\nx}}
\ncd{\BYY}{\frac{\partial\BY}{\partial\ny}}
\ncd{\BXY}{\frac{\partial\BX}{\partial\ny}}
\ncd{\BYX}{\frac{\partial\BY}{\partial\nx}}
\ncd{\BXXY}{\frac{\partial\BX}{\partial\nxy}}
\ncd{\BYXY}{\frac{\partial\BY}{\partial\nxy}}
\ncd{\BXab}{\BX_{a b}}
\ncd{\Acalxab}{{\mathcal{A}}^\x_{a b}}
\ncd{\Acalxyab}{{\mathcal{A}}^{\x \y}_{a b}}
\ncd{\Bcalxab}{\BX_{a b}}
\ncd{\Xcalab}{\Xcal^{\x \y}_{a b}}
\ncd{\Ccc}{\mathcal{C}_{cc}}
\ncd{\Cent}{\mathcal{C}_{ent}}
\ncd{\CX}{\mathcal{C}^X}
\ncd{\CY}{\mathcal{C}^Y}
\ncd{\Ccal}{\mathcal{C}}
\ncd{\Xcal}{\mathcal{X}}
\ncd{\kbar}{\bar{k}}
\ncd{\kybrack}{\left[v^2(v_s^2-1)\sin^2\theta + (1-v_sv\cos\theta)^2\right]}
\ncd{\logder}[2]{\frac{\partial\ln #1}{\partial\ln n_{#2}}}
\ncd{\deriv}[2]{\frac{\partial #1}{\partial #2}}
\ncd{\kybrackorth}{[v^2(v_s^2-1)+1]}
\ncd{\EX}{\mathcal{E}_{X}}
\ncd{\EY}{\mathcal{E}_{Y}}
\ncd{\EXY}{\mathcal{E}_{XY}}
\ncd{\EXb}{\bar{\mathcal{E}}_{X}}
\ncd{\EYb}{\bar{\mathcal{E}}_{Y}}
\ncd{\EXYb}{\bar{\mathcal{E}}_{XY}}
\ncd{\abar}{\bar\alpha}
\ncd{\bbarX}{\bar{\beta}_X}
\ncd{\bbarY}{\bar{\beta}_Y}
\ncd{\dBXX}{\frac{\BX}{\nx}(c_x^2-1)}
\ncd{\dBYY}{\frac{\BY}{\ny}(c_y^2-1)}
\ncd{\dBXY}{\frac{\BX}{\nx}\alpha}
\ncd{\dBXXY}{\frac{\AXY}{\nx}\EX}
\ncd{\dBYXY}{\frac{\AXY}{\ny}\EY}
\ncd{\dAXYXY}{\frac{\AXY}{\nxy}\EXY}
\ncd{\BYb}{\bar{\mathcal{B}}^Y}
\ncd{\AXYb}{\bar{\mathcal{A}}^{XY}}
\ncd{\rhoc}{\check\rho}
\ncd{\pc}{\check{p}}
\ncd{\Me}{m_x^y}

\begin{document}

\title{Relativistic Two-stream Instability}

\author{L. Samuelsson}

\affiliation{Nordita, Roslagstullsbacken 23, 106 91 Stockholm, Sweden}

\author{C. S. Lopez-Monsalvo, N. Andersson}

\affiliation{School of Mathematics, University of Southampton
\\ Southampton SO17 1BJ, UK}

\author{G. L. Comer}

\affiliation{Department of Physics \& Center for Fluids at All Scales,
Saint Louis University, St.~Louis, MO, 63156-0907, USA}

\date{\today}

\begin{abstract}
We study the (local) propagation of plane waves in a relativistic,
non-dissipative, two-fluid system, allowing for a relative velocity in the
``background'' configuration. The main aim is to analyze relativistic
two-stream instability. \ This instability requires a relative
flow --- either across an interface or when two or more fluids interpenetrate
--- and can be triggered, for example, when one-dimensional plane-waves
appear to be left-moving with respect to one fluid, but right-moving with
respect to another. \ The dispersion relation of the two-fluid system is studied for
different two-fluid equations of state: (i) the ``free'' (where there is no
direct coupling between the fluid densities), (ii) coupled, and (iii)
entrained (where the fluid momenta are linear combinations of the velocities)
cases are considered in a frame-independent fashion (\eg no restriction to
the rest-frame of either fluid). \ As a by-product of our analysis we
determine the necessary conditions for a two-fluid system to be causal and
absolutely stable and establish a new constraint on the entrainment.
\end{abstract}

\maketitle


\section{Introduction}
\label{intro}

Newtonian physics is replete with examples of multi-fluid systems, such
as diffusion, ion flow, superfluid Helium, and plasma discharge from the Sun.
\ In fact, large characteristic scattering times between different components
is more the norm than the exception. \ This leads to physical situations
where the various components can move independently of each other, be it
across an interface or through interpenetration. \ In this context, even heat
conduction in systems where all the matter flows together is a two-fluid
problem, \ie there is a heat flux in addition to the matter flux. \ Perhaps
not as widely appreciated is that the relativistic regime has its own set of
multi-fluid scenarios: neutrino streaming during supernovae, superfluid
neutrons and superconducting protons in neutron stars, and heat flow in a
cosmological setting, to name but a few.

A key issue is that relativistic fluids must be causal, meaning
that sound speeds, say, must be less than that of light. \ For fluids, there
are two entry points for causality: the microscopic where particle-particle
interactions are tracked and the macroscopic where fluid elements (large
enough to contain many particles, but small enough to be point-like with
respect to the total system) are monitored. \ Presumably, a fully relativistic
treatment at the microscopic level would lead to a set of fluid coefficients
(describing the equation of state, dissipation, etc.)~that would already
behave appropriately at the macroscopic level. \ However, there is a practical
problem: Equation of state determinations are notoriously difficult. \ This
makes a general analysis of relativistic fluid dynamics prohibitive, if not
impossible. \ Fortunately, one can make progress by imposing causality ``from
above'' and absolute stability (\ie real sound speeds) ``from below'' to
constrain the fluid coefficients.

In this paper, we will do this by analyzing the local propagation of plane
waves on a given (arbitrary) background spacetime. \ Compared to the standard
single-fluid analysis, we have more fluid degrees of freedom and need to
allow for relative flows between the various fluids. \ This is an essential
requirement for two-stream instability. \ Such instabilities are known to
exist for a variety of configurations. \ For shearing motion at an interface,
it is an example of Kelvin-Helmholtz instability. \ However, as far as we are
aware, generic two-stream instability has not been discussed previously in
relativity.

The two-stream instability has been well-documented for plasmas (where
it is known as the ``Farley-Buneman'' instability
\cite{farley63:_2stream,buneman59:_2stream}, see \cite{chen_plasma}
for a text-book discussion). It has also been suggested as the
mechanism behind star formation when two galaxies (whose angular
velocities are more or less anti-aligned) merge
\cite{lovelace1997:_gal_two_strm}. \ In the general relativistic
context, Chandrasekhar, Friedman, and Schutz (CFS)
\cite{chandrasekhar70:_grav_instab,friedman78:_secul_instab} have demonstrated
that oscillation modes in rotating, perfect fluids can become
two-stream unstable due to the emission of gravitational radiation. \
Here, the two ``fluids'' are the rotating mass, and the asymptotically
flat spacetime in which the fluid is embedded. \ Most recently, a
two-stream instability for superfluids has been proposed, with a
natural extension to a mixture of superfluid neutrons and
superconducting protons in neutron star cores
\cite{acp03:_twostream_prl,andersson04:_twostream}. \ Very recent results
suggest that this instability may act as a trigger mechanism for the enigmatic
spin glitches that have been observed in a number of radio pulsars
\cite{glampedakis09:_glitch_prl} (see also \cite{andersson04:_twostream} for
the first suggestion of a link between glitches and two-stream instability).

In what follows we will not restrict the discussion to any specific physical
system. Consequently, the analysis will be somewhat abstract. \ This
strategy can work because the two essential requirements for triggering 
two-stream instability is a relative flow between two fluids and a generic
interaction between them. \ This freedom to remain abstract illustrates the
general robustness of the instability and its presence in a diverse collection
of systems. \ Essentially, if the relative velocity is large enough that a
wave moves in one direction with respect to the rest-frame of one fluid, yet
the opposite direction with respect to the other fluid's rest-frame, then the
energy of the wave will be ``negative'' in one of the rest-frames and
therefore unbounded from below. \ Our main aim is to show that a causal and
absolutely stable system of two relativistic fluids can undergo two-stream
instability for a range of relative speeds and equation of state parameter
values.

The presentation of the results is organized as follows: Section \ref{formal}
recalls the multi-fluid formalism, and sets the stage for a plane-wave
analysis of the system. \ Section \ref{singwaves} considers sound waves for a
single fluid. \ The results are not new, but they  help  establish
basic techniques that carry over to the more complicated two-fluid
calculations discussed in Section \ref{twowaves}. \ The following
sub-sections consider three variations on the two-fluid equation of state:
(i) the ``free'' (where there is no direct coupling between the fluid
densities), (ii) coupled, and (iii) entrained cases. \ Finally, in Section
\ref{conclu}, we make our concluding remarks. \ Spacetime indices are denoted
by the first letters of the roman alphabet ($a, b, c$), constituent indices by
the last (x, y, z), and we adopt ``MTW'' (Misner, Thorne, and Wheeler
\cite{mtw73}) conventions for the metric signature.


\section{The Multi-fluid Formalism}
\label{formal}

We will use the approach to multi-fluid systems that was developed
originally by Carter \cite{carter89:_covar_theor_conduc} (see
\cite{andersson07:_livrev} for a recent review). \ In this description, the
main variables are the various fluxes (for particles and/or entropy),
to be denoted $n^a_\x$, and the equations of motion follow from a
suitably defined ``master'' function (\ie Lagrangian or equation of
state) $\Lambda$. \ In the single fluid case, $-\Lambda$ is equal to
the rest frame energy density $\rho$ .  \ We have here introduced the
convention of attaching a constituent index $\x$ to each
variable. This index is redundant for a single fluid, but necessary
when there are multiple fluids. \ The master function varies only with
the fluxes.
\ If the fluids are locally isotropic (\ie no preferred direction), as they
should be in the absence of anything else (such as an elastic solid), it is
clear that $\Lambda$ must be a function of only the various scalars that can
be formed from inner products of the fluxes.

We will focus on the case of two fluids (see \cite{andersson07:_livrev} for a
complete description), even though most of the equations in the general
discussion will carry enough constituent indices ($\x$, $\y$, etc.) to be valid for
any number of fluids. \ In the case of two components, the master function
depends on two distinct particle fluxes $n^a_\x$ and $n^a_\y$ and has the
functional dependence~\footnote{It is worth making the following remark on
the notation. \ Throughout the paper we only consider two fluids. \ They are
generally labelled by $\x$ and $\y$. \ However, in order to be economic in the
presentation we often treat the constituent index as abstract, meaning that
it can be either $\x$ or $\y$. \ That is, an equation written down explicitly
for fluid $\x$ takes exactly the same form for the other fluid once the index
$\x$ is replaced by $\y$ (and vice versa). \ We are aware that this
convention may be confusing at first, but it makes sense. \ Especially if one
wants to account for additional fluid components. \ Most of our equations
still remain valid in that case, although in each equation for fluid $\x$ one
has to sum over all the other fluids (i.e.~the sum runs over $\y \neq \x$).}
\beq
    \Lambda = \Lambda(\nx,\ny,\nxy) \ , \label{mf2}
\eeq
where $\nxy = - g_{a b} n^a_\x n^b_\y$. \ Note that constituent indices are
not summed over when repeated. \ As a matter of convenience repeated indices
are written only once; that is, we write $\nx$ (which is the squared particle
number density of the $\x^{\rm th}$-fluid) for $n^2_{\x\x}$ and so on.

The equations of motion become most transparent when expressed in terms of
the momentum $\mu^\x_a$ which is canonically conjugate to $n_\x^a$:
\beq
    \mu_a^\x = \BX n^\x_a +  \AXY n^\y_a \ , \label{mom}
\eeq
where
\beq
    \BX \equiv -2 \frac{\partial \Lambda}{\partial \nx}
        \quad , \quad
    \AXY \equiv -\frac{\partial \Lambda}{\partial \nxy}
               \ . \label{bxaxydef}
\eeq
Note that we have simplified the notation by not indicating explicitly
which variables are fixed when partial derivatives are taken. (The
functional dependence of the master function is clear from
\eqref{mf2}.)
\ From \eqref{mom} we see that the momentum $\mu^\x_a$ is not simply
proportional to its canonical conjugate $n^a_\x$, but is rather a linear
combination of all the fluxes. \ This is a result of the so-called
entrainment effect (see \cite{andreev75:_three_velocity_hydro} for an example
in superfluid Helium mixtures,
\cite{borumand96:_superfl_neutr_star_matter,comer03:_rel_ent,chamel06:_ent_cold_ns}
for relativistic, nuclear matter, or \cite{andersson08:_ent_ent} for a
treatment of entropy/matter entrainment and its importance for heat flow).

It is  convenient at this point to introduce a shorthand notation for
derivatives of these coefficients; namely,
\bea
    \Ccc^2 &\equiv& \frac{1}{\BX \BY} \left(2 n_\x n_\y
                 \frac{\partial \BX}{\partial \ny} \right)
                 \ , \cr
    \BX_{, \x \y} &\equiv& n_\x n_\y \frac{\partial \BX}{\partial \nxy}
                  \ , \cr
    \AXY_{, \x \y} &\equiv& n_\x n_\y \frac{\partial \AXY}{\partial
                   \nxy} \ . \label{twoderivs}
\eea
For the same reason we define the ``speed-of-sound'' $c^2_\x$ of the
$\x^{\rm th}$-fluid as
\beq
    c^2_\x = \frac{\partial \log \BX}{\partial \log n_\x} + 1 \ . \label{cs1}
\eeq
Finally, we introduce the ``perp'' operator
\beq
    \perp^{\x b}_a = \delta_a{}^b + u^\x_a u^b_\x
         \quad , \quad
    \perp^{\x b}_a u^a_\x = 0 \ , \label{perp}
\eeq
which can be used to construct, say, vectors that are orthogonal to $u^a_\x$.

We have chosen the fluxes $n^a_\x$ as our primary fields. \ However, there
is no reason why the momenta $\mu^\x_a$ could not be similarly adopted
\footnote{In fact, the celebrated Landau model for superfluid Helium mixes a
momentum with a flux, see for example, the discussion in
\cite{carter94:_canon_formul_newton_superfl}.}. \ This implies that the
mapping from one set of fields to the other must have an inverse. \ That is,
we see from \eqref{mom} that
\beq
 \left[\begin{array}{c} \mu^\x_a \\ \mu^\y_a \end{array}\right]
    = \left[\begin{array}{cc}
    \BX &  \AXY \\
    \AXY & \BY
    \end{array}\right]
    \left[\begin{array}{c}
    n^\x_a \\ n^\y_a
    \end{array}\right]
    \ , \label{momfluxmat}
\eeq
and thus $\BX \BY - ({\AXY})^2 \neq 0$. \ This will be a useful constraint
later, when we discuss the impact of entrainment on sound modes.

For the purely variational case (\ie no dissipation, imposed constraints,
etc.), the individual constituents are conserved
\cite{andersson07:_livrev,carter06:_crust}, so that we have for each component
\beq
    \nabla_a n_\x^a = 0 \ .
\eeq
The remaining equations of motion take the form
\beq
    n^a_\x \omega^\x_{a b} = 0 \ ,  \label{forceeqns}
\eeq
where the vorticity tensor $\omega^\x_{a b}$ is given by
\beq
    \omega^\x_{a b} \equiv 2 \nabla_{[a}\mu^\x_{b]} \ . \label{vort}
\eeq
As discussed, \eg, in \cite{andersson07:_livrev}, in the single-fluid case
these equations contain the same information as the standard set obtained
from the vanishing of the covariant divergence of the stress-energy-momentum
tensor. \ Eq.~\eqref{forceeqns} illustrates the geometrical significance of
the Euler equation as an integrability condition on the vorticity; \ie that
the particle flux nowhere pierces the surfaces defined by the two-form
$\omega^\x_{a b}$.

Below we will be analyzing plane-wave propagation on backgrounds such that
$\omega^\x_{a b} = 0$, the various background quantities are taken to be
constant, and there is a relative flow between the fluids. \ This implies a
linearization of the equations of motion; \ie
\beq
    \nabla_a \delta n^a_\x = 0
         \quad , \quad
    n_\x^a \nabla_{[a}\delta \mu^\x_{b]} = 0 \ . \label{pereqsx}
\eeq
Because there are several fluids, the variation $\delta \mu^\x_a$ is
significantly more complicated than, say, in the case of the standard perfect
fluid. \ In addition to individual bulk effects for each fluid, there can
also be cross-constituent effects due to coupling between the fluids. \ We
also have to consider the  entrainment.

Following \cite{andersson07:_livrev}, we can isolate the various effects in
the variation and write
\beq
    \delta \mu_a^\x = \left(\BXab + \Acalxab\right) \delta n^b_\x +
                      \left(\Xcalab + \Acalxyab\right) \delta n^b_\y \ ,
                      \label{muvarx}
\eeq
where the bulk effects are captured by
\beq
    \BXab = \BX \left(\perp^\x_{a b} - c^2_\x u^\x_a u^\x_b\right)
            \ , \label{bulk}
\eeq
the cross-constituent coupling through
\beq
    \Xcalab = - \Ccc \sqrt{\BX \BY} u^\x_a u^\y_b \ , \label{crosscoup}
\eeq
and the entrainment via the terms
\bea
    \Acalxab &=& - \left[\BX_{, \x \y} \left(u^\x_a u^\y_b + u^\x_b u^\y_a
                 \right) + \frac{n_\y}{n_\x} \AXY_{, \x \y} u^\y_a u^\y_b
                 \right] \ , \\
              && \cr
    \Acalxyab &=& \AXY \!\!\perp^\x_{a b} - \left[\left(\AXY + \frac{n_\x}{n_\y}
                  \BX_{, \x \y}\right) u^\x_a u^\x_b + \frac{n_\y}{n_\x}
                  \BY_{, \x \y} u^\y_a u^\y_b + \AXY_{, \x \y} u^\y_a
                  u^\x_b\right] \ . \label{entraincoup}
\eea
In these expressions the flux $n^a_\x$ has been decomposed as $n^a_\x = n_\x
u^a_\x$, where $u^a_\x$ is the unit ($u^\x_a u^a_\x = - 1$) four-velocity of
the $\x$-fluid elements.

\section{Single-fluid Sound Waves}
\label{singwaves}

To set the stage for the general analysis it is useful to first consider the
nature of sound waves in a single, perfect fluid. \ To do this, we perform a
local analysis of linear perturbations of the fluid (keeping the metric fixed)
on a generic background. \ In particular, plane-wave propagation corresponds
to the Ansatz
\beq
    \delta n^a_\x  = A^a_\x \exp(i k_b x^b) \ , \label{plwave}
\eeq
where the amplitude $A^a_\x$ and wave four-vector $k^a$ have vanishing
covariant derivatives. \ Recall that we assume all unperturbed quantities to
be similarly constant. \ In particular, the background vorticity simply
vanishes. \ From \eqref{muvarx} above we have
\beq
    \delta \mu^\x_a = \Bcalxab \delta n^b_\x \ . \label{varmu1}
\eeq
Note that we have continued to use a constituent index, even though we are
dealing with a single fluid. \ This allows for some economy of presentation
since many of the formulas will apply later, except that the index $\x$ will
then range over two fluids. \ Of course, waves in a system are such that the
constant wave vector $k_a$ is the same for all the fluids. \ Hence, it does
not carry a constituent index.

For convenience we will work in the material frame associated with the fluid.
\ This means that $k_a$ and $A^a_\x$ will each be written as two pieces by
utilizing the ``perp'' operator introduced in \eqref{perp}:
\beq
    k_a = k_\x \left(\sigma_\x u^\x_a + \hat{k}^\x_a\right) \ , \label{kdef}
\eeq
where $\sigma_\x$ and the wave vector $k^\x_a$ (with magnitude $k_\x$)
are
\beq
    k_\x \sigma_\x = - k_a u^a_\x
                \quad , \quad
    k^\x_a = \perp^b_{\x a} k_b \equiv k_\x \hat{k}^\x_a \ .
\eeq
Similarly, we can decompose the wave amplitude as
\beq
    A^a_\x = A^\x_{||} u^a_\x + A_{\x \perp}^a \ ,
\eeq
where
\beq
    A^\x_{||} = - u^\x_a A^a_\x \quad , \quad A_{\x \perp}^a = \perp^a_{\x b}
                A^b_\x \ .
\eeq
Note that $\sigma_\x$ and $\hat{k}_\x^a$ are measured by an observer moving
with the fluid. \ It will be obvious from the dispersion relation constructed
below that $\sigma_\x$ measures the phase velocity of the waves as seen in
the fluid frame. \ Furthermore, it is easy to show that evaluating
$\sigma_\x$ in a frame moving relative to the fluid leads to the standard
Lorentz transformation of velocities.

With these preliminaries, the perturbation equations \eqref{pereqsx} reduce to
\bea
     0 &=& k_a A^a_\x \ , \label{cons1} \\
     0 &=& n^a_\x k_{[a} \BX_{b] c} A^c_\x \ . \label{euler1}
\eea
The first of these relations shows that the waves are transverse in
the spacetime sense. \ The dispersion relation can be easily obtained
by contracting the second equation with $k^b$. (In the more
complicated two-fluid analysis below, we will in general have to
consider the vanishing of a $4 \times 4$ determinant.) \ Assuming that
$n^\x_c A^c_\x \neq 0$, and after several steps of algebra, the
dispersion relation reduces to
\beq
    \sigma^2_\x - c^2_\x = 0 \ . \label{disper1}
\eeq
In our homogeneous plane-wave setting it is clear that the group and phase
velocities coincide so that we can introduce the speed of sound in the
standard way as $c^2_\x = \sigma^2_\x$.

To see that this is equivalent to the usual single-fluid result (as in,
say, \cite{weinberg72:_book}), it will suffice to introduce the pressure and
recall that $\rho = - \Lambda$ in the single fluid case. \ From the
definition above for $\BX$, cf.~\eqref{bxaxydef}, we see that
\beq
    {\rm d} \rho = \BX n_\x {\rm d} n_\x \ .
\eeq
Moreover, the  pressure $p$ is defined by the standard thermodynamic relation, such that
\beq
    p = - \rho + n_\x{ { \rm d} \rho \over {\rm d} n_\x} =  - \rho + \BX n^2_\x \ .
\eeq
This implies
\beq
    {\rm d} p = \left(1 + \frac{\partial \log \BX}{\partial \log n_\x}\right)
                {\rm d} \rho \ , 
\eeq
and we have
\beq
    \frac{{\rm d} p}{{\rm d} \rho} = 1 + \frac{\partial \log \BX}{\partial
       \log n_\x} = c^2_\x \ ,
\eeq
where ${\rm d} p/{\rm d} \rho$ is the usual form of the sound speed squared.

In order to pave the way for the more complicated multi-fluid case to be
discussed below it is useful to examine the properties of the various vectors
we have introduced. \ Starting with the wave vector $k^a$ we see that
\beq
    k_a k^a = k_\x^2 \left(1 - c_\x^2\right) \ . \label{kmag}
\eeq
Thus, for causal wave propagation ($c_\x^2 \leq 1$), $k^a$ is spacelike. \
For the wave amplitude we find that, when the force equation
\eqref{euler1} is evaluated in terms of the solution to the
dispersion relation,
\beq
    A_{\x \perp}^a = \sigma_\x A^\x_{||} \hat{k}_\x^a \ .
\eeq
The waves are therefore longitudinal in the normal, three-dimensional sense. \
On the other hand, the transverse nature \eqref{cons1} of the waves in the
four-dimensional sense implies that
\beq
  A^2_\x = A_{\x \perp}^2 \left(1 - c_\x^{-2}\right)
\eeq
so that $A^a_\x$ is timelike (and we can choose it to be future
pointing) for causal waves. \ Note that since the flux $n^a_\x$ is
also timelike this implies that $n^a_\x A^\x_a < 0$ and thus the
degenerate case of \eqref{disper1} is ruled out by causality.

Before we conclude this section, it is useful to consider what constraints
\eqref{cs1} imposes on the equation of state. \ First of all, the causality
requirement leads immediately to
\beq
    \frac{\partial \log \BX}{\partial \log n_\x} + 1 \le 1 \ .
\eeq
In addition, we must have $c_\x^2 \ge 0$ in order to avoid absolute
instabilities (complex wave speeds). This implies
\beq
    \frac{\partial \log \BX}{\partial \log n_\x} + 1 \ge 0 \ . \label{absstab1}
\end{equation}
Combining the two results we see that we must have
\beq
    - 1 \le \frac{\partial \log \BX}{\partial \log n_\x} \le 0 \ .
    \label{stabcaus1}
\eeq
In the next section we will extend this type of analysis to the two-fluid
model.


\section{Sound Waves for the General Two-fluid System}
\label{twowaves}

We want to work out the  dispersion relation for wave propagation
in a two-fluid system, using \eqref{plwave} as the starting point. \ An
important addition to the problem of plane waves is a new ``parameter'',
the relative flow between the two fluids. \ We will represent this flow
 by the relative velocity $v^a_{\x \y}$ of the $\y^{\rm th}$-fluid with
respect to the frame of the $\x^{\rm th}$-fluid:
\beq
    \gamma_{\x \y} v^a_{\x \y} = \perp^{\x a}_b u^b_\y \ , \label{vxy}
\eeq
where $v_{\x \y}$ represents the magnitude of the relative flow and
\beq
    \gamma_{\x \y} = \gamma_{\y \x} = - u^c_\x u^\y_c =
                     \frac{1}{\sqrt{1 - v^2_{\x \y}}} \ .
\eeq
This leads to
\beq
    u^a_\y = \gamma_{\x \y}\left(u^a_\x + v^a_{\x \y}\right) \ .
\eeq
The mode speed $\sigma_\x$ and wave (three-) vector $k^\x_a$ can be defined as
before. \ The insertion of $\gamma_{\x \y}$ into \eqref{vxy} makes the
$\x^{\rm th}$-fluid's proper time the standard for setting velocities. \ Since the
dispersion relation below is a scalar equation, we will have in several
places the inner product $\hat{v}^a_{\x \y} \hat{k}^\x_a$ (where
$\hat{v}^a_{\x \y} = v^a_{\x \y}/v_{\x \y}$). \ It is useful to
write this in terms of the angle $\theta_{\x \y}$ between the two vectors:
\beq
    \hat{v}^a_{\x \y} \hat{k}^\x_a = \cos \theta_{\x \y} \ .
\eeq

An important subtlety must be recognized, however: The three quantities
$\sigma_\x$, $k^\x_a$, and $v^a_{\x \y}$ are what would be measured by an
observer flowing  with the $\x^{\rm th}$-fluid. \ We could have equally as
well chosen the material frame attached to the other fluid (or some other
observer). \ As one might expect, there are well-defined transformations
between the two descriptions (which will be needed later). \ The relative flow
$v^a_{\y \x}$ of the $\x^{\rm th}$-fluid with respect to the
$\y^{\rm th}$-fluid frame is related to $v^a_{\x \y}$ via
\beq
    v^a_{\y \x} = - \gamma_{\x \y} \left(v^2_{\x \y} u^a_\x +
                  v^a_{\x \y}\right) \ ,
\eeq
where we have used $v_{\y \x} = v_{\x \y}$. \ It is also useful to note that
\bea
    u^a_\x &=& - v^{- 2}_{\x \y} \left(v^a_{\x \y} + \gamma^{- 1}_{\x \y}
               v^a_{\y \x}\right) \ , \cr
    u^a_\y &=& - v^{- 2}_{\x \y} \left(v^a_{\y \x} + \gamma^{- 1}_{\x \y}
               v^a_{\x \y}\right) \ . \label{423vel}
\eea
Because $k_a$ is not attached to either fluid frame, we must have
\beq
    k_a = k_\y \left(\sigma_\y u^\y_a + \hat{k}^\y_a\right) = k_\x
         \left(\sigma_\x u^\x_a + \hat{k}^\x_a\right) \ . \label{kequal}
\eeq

By contracting each four-velocity in \eqref{423vel} with the wave-vector
$k_a$, we obtain a matrix equation for $[k_\x~k_\y]^{\rm T}$; namely,
\beq
    \left[\begin{array}{cc}
    v_{\x \y} \sigma_\x - \cos \theta_{\x \y} &  - \gamma^{- 1}_{\x \y}
    \cos \theta_{\y \x} \\
    - \gamma^{- 1}_{\x \y} \cos \theta_{\x \y} &
    v_{\x \y} \sigma_\y - \cos \theta_{\y \x}
    \end{array}\right]
    \left[\begin{array}{c}
    k_\x \\
    k_\y
    \end{array}\right] =
    \left[\begin{array}{c} 0 \\ 0 \end{array}\right]
    \ .
\eeq
Obviously the determinant of the $2 \times 2$ matrix must vanish. \ This
leads to
\beq
    \sigma_\y = \cos \theta_{\y \x} \frac{\sigma_\x - v_{\x \y} \cos
    \theta_{\x \y}}{v_{\x \y} \sigma_\x - \cos \theta_{\x \y}} \ .
    \label{sigtrans}
\eeq
It is not difficult to show that if $\sigma^2_\x \le 1$ then $\sigma^2_\y \le
1$. This is natural given that causality is a frame-independent requirement.

Meanwhile, the equation of flux conservation is the same as \eqref{cons1}
(except $\x$ ranges over two values). \ The conservation of vorticity
equations become
\bea
    0 &=& K^\x_{a b} A^b_\x + K^{\x \y}_{a b} A^b_\y \ , \cr
    0 &=& K^\y_{a b} A^b_\y + K^{\y \x}_{a b} A^b_\x \ , \label{pereqns2}
\eea
where the ``dispersion'' tensors are
\bea
    K^\x_{a b} &=& n^c_\x \left(k_{[c}\BX_{a]b} + k_{[c}\mathcal{A}^\x_{a]b}
                   \right) \ , \cr
    K^{\x \y}_{a b} &=& n^c_\x \left(k_{[c} \Xcal^{\x \y}_{a]b} +
                        k_{[c}\mathcal{A}^{\x \y}_{a]b}\right) \ .
                        \label{disptens}
\eea
Note that $K^\y_{a b}$ and $K^{\y \x}_{a b}$ are obtained via the interchange
of $\x \leftrightarrow \y$ in equation \eqref{disptens}.

In order to solve \eqref{pereqns2}, we obviously need the four inverses
\beq
    \tilde{K}^{a c}_\x K^\x_{c b} = \delta^a{}_c
         \quad , \quad
    \tilde{K}^{a c}_{\y \x}  K^{\x \y}_{c b} = \delta^a{}_c \ ,
\eeq
to exist---\ie the determinants of $K^\x_{a b}$ and $K^{\x \y}_{a b}$ do not
vanish---so that we can write
\beq
    0 = \left(\tilde{K}^{a c}_\y K^{\y \x}_{c b} - \tilde{K}^{a c}_{\y \x}
        K^\x_{c b}\right) A^b_\x \equiv {\cal M}_{a b} A^b_\x \ .
\eeq
The only way to get a non-trivial solution is to have a $k_a$ such that
\beq
    \epsilon^{a_1 a_2 a_3 a_4} \epsilon^{b_1 b_2 b_3 b_4} {\cal M}_{a_1 b_1}
    {\cal M}_{a_2 b_2} {\cal M}_{a_3 b_3} {\cal M}_{a_4 b_4} = 0 \ .
    \label{gensol}
\eeq

Written out in full \eqref{gensol} is a quite busy expression. \ This should
come as no surprise since the two-fluid problem is significantly more
complicated than a single fluid having bulk contributions coming from
$\BX_{a b}$, and the cross-coupling from $\Xcal^{\x \y}_{a b}$ (in the case
of two, co-moving constituents \cite{andersson07:_livrev}). \ For two-fluid
systems, the two constituents move independently and there are the additional
contributions $\mathcal{A}^\x_{a b}$ and $\mathcal{A}^{\x \y}_{a b}$ coming
from entrainment.

In order to simplify the problem, it is convenient to isolate further the
different contributions that appear in the dispersion matrices. \ The bulk
contribution in $K^\x_{a b}$ can be reduced to
\bea
    b^\x_{a b} &=& n^c_\x k_{[c}\BX_{a]b} \cr
               &=& - \frac{1}{2} \BX n_\x \left(k_\x \sigma_\x \perp^\x_{a b}
                   + c^2_\x k^\x_a u^\x_b\right) \ , \label{bulkmat}
\eea
while its entrainment piece becomes
\bea
    a^\x_{a b} &=& n^c_\x k_{[c}\mathcal{A}^\x_{a]b} \cr
               &=& \frac{1}{2} \gamma_{\x \y} n_\x \Bigl\{
                   \mathcal{B}^\x_{, \x \y} \left[\left(k_\x \sigma_\x
                   v^{\x \y}_a - 2 k^\x_a\right) u^\x_b - k^\x_a
                   v^{\x \y}_b\right]  \cr
                && + \gamma_{\x \y} \frac{n_\y}{n_\x}
                   \mathcal{A}^{\x \y}_{, \x \y} \left(k_\x \sigma_\x
                   v^{\x \y}_a - k^\x_a\right) \left(u^\x_b + v^{\x \y}_b
                   \right)\Bigr\} \ . \label{entmat1}
\eea
Meanwhile, the cross-coupling in $K^{\x \y}_{a b}$ is simply given by
\bea
    x^{\x \y}_{a b} &=& n^c_\x k_{[c} \Xcal^{\x \y}_{a]b} \cr
                    &=& - \frac{1}{2} \mathcal{C}_{cc} \gamma_{\x \y} n_\x
      \sqrt{\BX \BY} k^\x_a \left(u^\x_b + v^{\x \y}_b\right) \ , \label{ccmat}
\eea
but the entrainment has significantly more presence:
\bea
    a^{\x \y}_{a b} &=& n^c_\x k_{[c}\mathcal{A}^{\x \y}_{a]b} \cr
    &=& \frac{n_\x}{2} \left\{- \AXY k_\x \sigma_\x
    \perp^\x_{a b} - \left[\AXY + \frac{n_\x}{n_\y} \mathcal{B}^\x_{, \x \y} +
    \gamma_{\x \y} \left(\gamma_{\x \y} \frac{n_\y}{n_\x}
    \mathcal{B}^\y_{, \x \y} + \mathcal{A}^{\x \y}_{, \x \y}\right)\right]
    k^\x_a u^\x_b \right. \cr
    && \left. + \gamma_{\x \y} k_\x \sigma_\x \left(\gamma_{\x \y}
    \frac{n_\y}{n_\x} \mathcal{B}^\y_{, \x \y} +
    \mathcal{A}^{\x \y}_{, \x \y}\right) v^{\x \y}_a u^\x_b\right\} \ .
    \label{entmat2}
\eea

A multi-fluid system must have non-zero bulk properties (unless a fluid
vanishes completely). \ The other terms can be absent, depending on the
equation of state. \ In what follows we will systematically increase the
complexity by considering in turn the different components of two-fluid
physics.

\subsection{Dispersion Relation: Free Case}
\label{DR-free}

Let us first consider the case of two completely uncoupled fluids. \ Then we
have only $b^\x_{a b}$ non-zero. \ In lieu of \eqref{gensol}, it is easier to
get the dispersion relation by contracting the free indices in
\eqref{pereqns2} with $k_a$. \ This results in the simple $2 \times 2$ matrix
problem
\beq
    \left[\begin{array}{cc}
    \BX \left(\sigma^2_\x - c^2_\x\right) &  0 \\
    0 & \BY \left(\sigma^2_\y - c^2_\y\right)
    \end{array}\right]
    \left[\begin{array}{c}
    u^\x_a A^a_\x \\
    u^\y_a A^a_\y
    \end{array}\right] =
    \left[\begin{array}{c} 0 \\ 0 \end{array}\right]
    \ , \label{bulkalg}
\eeq
and the dispersion relation is simply
\beq
    \left(\sigma^2_\x - c^2_\x\right) \left(\sigma^2_\y - c^2_\y\right) = 0
    \ . \label{freedisp}
\eeq
By construction $\sigma^2_\x$  is the squared phase (three-) velocity as
measured in the $\x^{\rm th}$-fluid frame (similarly for $\sigma^2_\y$).

The outcome of this analysis is that the dispersion relation \eqref{freedisp}
allows four non-trivial solutions consisting of the roots of
\beq
    \sigma^2_\x = c_\x^2
\eeq
for any $\x$. \ These roots correspond to  $\pm c_\x$ evaluated in the
$\x^{\rm th}$-fluid frame and are just Lorentz transformed if evaluated in
another frame. \ Thus, as expected in the case of zero coupling between the
fluids, the quantity $c_\x$ can be interpreted as the sound velocity of the
$\x^{\rm th}$-fluid as measured in its own (background) rest frame. \ This is
the obvious generalisation of the single fluid result. \ It follows that if
$c_\x$ is subluminal in its own rest-frame it is so in all other frames as
well. \ Also, since absolute instability should be evaluated at zero relative
velocity it is clear that the constraint $0 \le c^2_\x \le 1$ remains as a
condition for the master function.

The main conclusion from this discussion is that the constraints on the
equation of state are easily generalized to the uncoupled two-fluid
problem. \ We must thus require that the equation of state satisfy
\eqref{stabcaus1} for both fluids in order for the system to be absolutely
stable and give rise to causal wave propagation. \ There are no dynamical
instabilities present in this case. \ In what follows, we caution that the
$c_\x$ can be understood as ``sound'' speeds only in this completely free
case. \ When  fluid couplings are operative, the phase velocities will no
longer simply equal these free sound speeds. \ But in order to make progress,
we will continue to impose \eqref{stabcaus1} throughout.

\subsection{Dispersion Relation: Cross-constituent Coupling}
\label{DR-coup}

We now allow for $x^{\x \y}_{a b}$ to be non-zero. \ From \eqref{ccmat}, we
see that this will introduce $\Ccc$ in addition to $\BX$ and $c^2_\x$. \
Using the same  contractions with $k_a$ as in the free case, we again
get a $2 \times 2$ matrix problem; \ie
\beq
    \left[\begin{array}{cc}
    \BX \left(\sigma^2_\x - c^2_\x\right) & - \sqrt{\BX \BY} \Ccc \\
    - \sqrt{\BX \BY} \Ccc & \BY \left(\sigma^2_\y - c^2_\y\right)
    \end{array}\right]
    \left[\begin{array}{c}
    u^\x_a A^a_\x \\
    u^\y_a A^a_\y
    \end{array}\right] =
    \left[\begin{array}{c} 0 \\ 0 \end{array}\right]
    \ . \label{ccalg}
\eeq
The dispersion relation is now
\beq
    \left(\sigma_\x^2 - c_\x^2\right) \left(\sigma^2_\y - c^2_\y\right) =
    \Ccc^2 \ . \label{dispcc}
\eeq
In order for $\Ccc^2$ to be less than zero, the equation of state would have
to allow either of the $\BX$ to be negative. \ For ordinary matter, or
entropy, this is not generally the case. \ Hence, we do not consider this
possibility here. \ We also need to point out that \eqref{sigtrans} must be
used to get a dispersion relation solely in terms of $\sigma_\x$. \ Clearly,
the cross-coupled case is more complicated than the free problem. \ However, it is also much more
interesting and relevant. \ As we will soon see, the richer phenomenology
allows for two-stream instability.

Addressing first the question of absolute stability we set $v_{\x \y} = 0$ to
find
\beq
    \sigma^2_\x = \frac{1}{2} \left(c^2_\x + c^2_\y \pm \sqrt{\left(c^2_\x -
                c^2_\y\right)^2 + 4 \Ccc^2}\right) \ . \label{ccsig}
\eeq
In order to avoid complex $\sigma^2_\x$, the discriminant of \eqref{ccsig}
must be positive, which is evident for $\Ccc^2 \ge 0$. \ Recall that absolute
stability means $\sigma^2_\x \ge 0$. \ Clearly, it is sufficient to require
that
\beq
    \left(c^2_\x + c^2_\y\right)^2 \ge \left(c^2_\x - c^2_\y\right)^2 + 4
          \Ccc^2 \ .
\eeq
The second term of \eqref{ccsig} will always be less than the
first if
\beq
    c^2_\x c^2_\y \ge \Ccc^2 \ . \label{ccstab}
\eeq
In other words, since we expect to have $\BX \BY > 0$, absolute stability
constrains the equation of state to satisfy
\beq
  \frac{\partial\log\mathcal{B}^\x}{\partial\log n_\y}
  \frac{\partial\log\mathcal{B}^\y}{\partial\log n_\x} \le
  \left(1 + \frac{\partial\log\mathcal{B}^\x}{\partial\log n_\x}\right)
  \left(1 + \frac{\partial\log\mathcal{B}^\y}{\partial\log n_\y}\right) \ .
\eeq

The causality constraint requires $\sigma^2_\x \le 1$. \ For $\Ccc^2 \ge 0$
and satisfying \eqref{ccstab}, we need only make the ``+'' solution in
\eqref{ccsig} causal, since the ``$-$'' solution is always smaller. \ We find
that causality is ensured if
\beq
    \Ccc^2 \le \left(1 - c^2_\x\right) \left(1 - c^2_\y\right) \ ,
               \label{cccaus}
\eeq
which, in terms of the equation of state, translates into
\beq
  \frac{\partial\log\mathcal{B}^\x}{\partial\log n_\y}
  \frac{\partial\log\mathcal{B}^\y}{\partial\log n_\x} \le
  \frac{\partial\log\mathcal{B}^\x}{\partial\log n_\x}
  \frac{\partial\log\mathcal{B}^\y}{\partial\log n_\y} \ .
\eeq
Note that both \eqref{ccstab} and \eqref{cccaus} restrict $\Ccc^2$ from
above. \ In general, we can show that if $c_\x^2 + c_\y^2 \le 1$ then any
absolutely stable equation of state is also causal. \ Conversely, if $c_\x^2
+ c_\y^2 \ge 1$, any causal equation of state is absolutely stable.

Given an equation of state that is causal and absolutely stable, we can now
determine if a dynamical two-stream instability is present by solving the
dispersion relation \eqref{dispcc} for some relative flow (as parameterized by
$v_{\x \y}$). \ Writing down the general solution is not difficult, but it is
 instructive to first focus on the slow velocity limit. \ Assuming
that $v_{\x \y}$ and $\sigma_\x$ are both much smaller than unity (the speed
of light) in \eqref{sigtrans}, the dispersion relation \eqref{dispcc} becomes
\beq
    \left(\sigma^2_\x - c_\x^2\right) \left[\left(\sigma_\x - v_{\x \y}
           \cos \theta_{\x \y}\right)^2 - c^2_\y\right] = \Ccc^2 \ .
\eeq
Introducing  new variables
\beq
  x = \frac{\sigma_\x}{c_\y}
      \quad , \quad
  y = \frac{v_{\x \y} \cos \theta_{\x \y}}{c_\y}
      \quad , \quad
  b^2 = \left(\frac{c_\x}{c_\y}\right)^2
        \quad , \quad
  a^2 = \frac{\Ccc^2}{c^4_\y} \ , \label{rescale}
\eeq
we get
\beq
    \frac{x^2 - b^2}{a^2} \left[\left(x - y\right)^2 -1\right] = 1 \ .
\eeq
As one might have expected, the problem is now identical to the Newtonian plane-wave
problem discussed by Andersson, Comer, and Prix \cite{andersson04:_twostream}.
Hence, we can learn from their results. \ They demonstrate that a two-stream
instability may operate above a critical relative flow. \ Their particular
example corresponds to $a^2 = 0.0249$ and $b^2 = 0.0379$. \ For this case
they find an instability in the range $0.6 < y < 1.5$. \ This means that the
system becomes unstable for $c_\x y > 0.6$. \ This flow is clearly
sub-luminal as long as $c_\x < 1$, but one may suspect that the linear
approximation that we have used is not very accurate. \ Still, this is a
useful first demonstration that the two-stream instability will operate also
in relativistic systems.

Before we turn our attention to the full relativistic case it is useful to
check if the particular example used by Andersson et al.\
\cite{andersson04:_twostream} obeys the causality and absolute stability
criteria derived above. \ First we note that, due to the presence of a
velocity scale given by the speed of light, in relativity we cannot
completely scale out the velocities. \ Thus the relativistic analysis of
stability will in general contain an extra parameter compared to the
Newtonian case. \ Here we shall take that parameter to be $c_\y$ which,
without loss of generality, can be taken to be larger than $c_\x$. \ Using
these parameters the absolute stability criterion \eqref{ccstab} becomes just
\beq
    \frac{a^2}{b^2} \le 1 \ ,
\eeq
which is satisfied in the model discussed above. \ The causality condition
\eqref{stabcaus1} enters only indirectly as we have been able to re-scale in
terms of $c_\y$ and thereby get dimensionless variables. \ We conclude that
the Newtonian model of Andersson et al.\ is reasonable also from this
perspective as long as $c_\y$ is not very close to the speed of light.

We now turn  to the relativistic dispersion relation \eqref{ccsig}. \ Written
as an equation for $\sigma_\x$ it constitutes a non-trivial quartic. \ If
we use the same re-scalings as in \eqref{rescale}, then \eqref{ccsig} becomes
\beq
   \frac{x^2 - b^2}{a^2} \left[\frac{\left(x - y\right)^2}{\gamma^{-2}_{\x \y}
   \left(1 - c^2_\y x^2 \right) + c^2_\y \left(x - y\right)^2} - 1\right] =
   1 \ . \label{ccquartic}
\eeq
Some immediate insight is obtained by considering the ultra-relativistic
limit for the background flow, where $v_{\x \y} \to 1$ or
$\gamma^{- 2}_{\x \y} \to 0$, and the limit where the wave vector becomes
perpendicular to the background flow, \ie $\theta_{\x \y} \to \pi/2$. \ In
both limits the two-stream instability ceases to operate. 

In the ultra-relativisitic limit, the wave-speed tends to
\beq
    \sigma_\x \to \left\{\begin{array}{l}
                    \pm\sqrt{c^2_\x + \frac{\Ccc^2}{1 - c^2_\y}} \\
                    \cos \theta_{\x \y} \quad \mbox{(double root)}
                    \end{array}\right. \ . \label{ultrarel}
\eeq
If the propagation is to remain causal, we must have $\Ccc^2 \to 0$ as $c_\y
\to 1$. \ Also there is no two-stream instability since $\Ccc^2 \ge 0$. \ This
might seem surprising, since a two-stream instability requires a ``window'' of
background flows for modes to appear, say, left-moving in one frame but
right-moving in the other. \ But as $v_{\x \y} \to 1$ the relative flow is at
its maximum, and yet the instability window is closed. \ In fact, this
behaviour was seen already by Andersson et~al.~\cite{andersson04:_twostream}
in the Newtonian regime. \ From \eqref{rescale} we also see that $y \to 0$
as $\theta_{\x \y} \to \pi/2$. \ This turns \eqref{ccquartic} into a
quadratic for $x^2$, and one finds that the discriminant is positive for the
range of values for $\Ccc^2$ that yield absolute stability and causality. \
Obviously, $y$ is the effective ``window'' of the background flow and it is
completely closed for $\theta_{\x \y} = \pi/2$. 

\begin{figure}
  \includegraphics[height=0.6\textwidth,width=0.9\textwidth,clip]
   {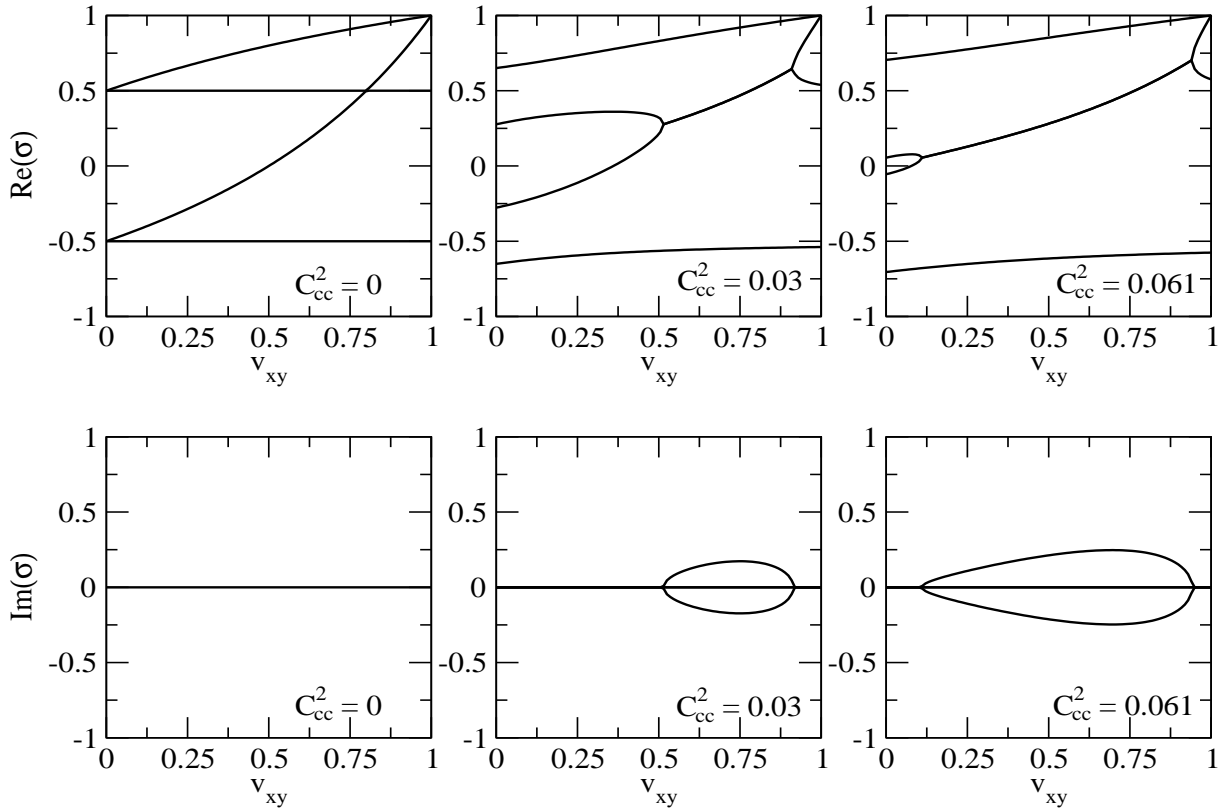}
   \caption{Plots of the real (top) and imaginary (bottom) parts of the mode
   frequencies $\sigma_\x$ as functions of $v_{\x \y}$, for $c_y = 0.5$, $b^2
   = 1.0$, and $\theta_{\x \y} = 0$. \ The merger of two frequencies, and
   subsequent non-zero imaginary values, signal the presence of a two-stream
   instability.} \label{sigcccplt0}
\end{figure}

Although the general solution to \eqref{ccquartic} is readily availiable, it
is quite complicated and offers very little additional insight. \ Instead of
writing it down we will tackle the problem numerically. \ The parameter
values are restricted to those that maintain absolute stability and
causality. \ Figure~\ref{sigcccplt0} provides plots of the real and imaginary
parts for the four solutions to \eqref{ccquartic} in the aligned case. \ The
solutions for $\sigma_\x$ are taken to be functions of the relative flow
parameter $y$ and the coupling $\Ccc^2$, with $c_\y = 0.5$, $b^2 = 1$, and
$\theta_{\x \y} = 0$. \ Non-zero values for ${\rm Im}\sigma_\x$ in the
figures indicate the presence of an unstable mode. \ The appearance of
unstable modes is reflected in the real parts wherever two frequencies merge.
\ This behaviour is typical for this kind of dynamical instability. \ The
results for misaligned flows, with $\theta_{\x \y} \neq 0$ are very similar
to those shown in Figure~\ref{sigcccplt0}. \ As $\theta_{\x \y}$ increases
the $[y,\Ccc^2]$ region of instability moves towards higher relative
velocities, eventually leading to regions that stretch essentially all the
way to $v_{\x \y} = 1$.

\subsection{Dispersion Relation: Aligned or Anti-aligned Flows}
\label{DRalign}

Now that we have established the presence of the two-stream instability for
arbitrary background flows, we will consider the more restricted case of
aligned or anti-aligned background flows. \ This will simplify the dispersion
relation so that analytical insight can be more easily acquired. \ It also
reduces the parameter space, thus allowing a more focused numerical analysis.

By aligned or anti-aligned flow we mean that the wave propagation is
aligned (or anti-aligned) with the relative velocity of the two fluids;
specifically, $\theta_{\x \y} = (0,\pi)$ so that
\beq
    \hat{k}^\x_a = \epsilon_{\x \y} \hat{v}^{\x \y}_a
                   \quad , \quad
    \epsilon_{\x \y} \equiv \cos \theta_{\x \y} = \pm 1 \ .
\eeq
This leads naturally to the statement that the wave vector is a linear
combination of the background flows:
\bea
    k_a &=& \frac{k_\x}{v_{\x \y}} \left[\left(v_{\x \y} \sigma_\x -
            \epsilon_{\x \y}\right) u^\x_a + \epsilon_{\x \y}
            \gamma^{- 1}_{\x \y} u^\y_a\right] \cr
        &=& \frac{k_\y}{v_{\x \y}} \left[\left(v_{\x \y} \sigma_\y -
            \epsilon_{\y \x}\right) u^\y_a + \epsilon_{\y \x}
            \gamma^{- 1}_{\x \y} u^\x_a\right] \ . \label{alignvecs}
\eea
Equating coefficients in \eqref{alignvecs} leads to
\beq
    k_a = \frac{1}{\gamma_{\x \y} v_{\x \y}} \left(k_\y \epsilon_{\y \x}
          u^\x_a + k_\x \epsilon_{\x \y} u^\y_a\right) \ . \label{kalign}
\eeq
Because of \eqref{kalign}, our original four-dimensional linear algebra
problem for $(u^\x_a A^a_\x,u^\x_a A^a_\y,u^\y_a A^a_\x,u^\y_a
A^a_\y)^\mathrm{T}$ has been reduced to a two-dimensional one for
$(u^\x_a A^a_\x,u^\y_a A^a_\y)^\mathrm{T}$.

\subsection{Dispersion Relation: Role of Entrainment}
\label{DR-entrain}

Up to this point we have introduced three equation of state parameters
($c_\x$, $c_\y$, and $\Ccc$) that are obtained as second derivatives of the
master function. \ When entrainment is included in the model we see from
\eqref{twoderivs} that we need two additional variables to describe the
general case. \ Given that we have established the two-stream instability for
general cross-constituent coupling and arbitrary background flow, the main
reason for discussing the role of the entrainment is to highlight the basic
feature that the instability can be triggered by a variety of interactions. \
We will simplify the entrainment case by assuming that the relative velocity
$v^a_{\x \y}$ is much smaller than the speed of light and that the flows are
aligned (in the sense of the previous section). \ This does not mean,
however, that the individual flows $u^a_\x$ have to be similarly restricted. 
Neither do the sound and wave speeds $c_\x$ and $\sigma_\x$ have to be small.

If we keep the relative flow to $\mathcal{O}(v^2_{\x \y})$, the master
function can be approximated as
\cite{andersson02:_oscil_GR_superfl_NS,andersson01:_dyn_superfl_ns}
\beq
    \Lambda = \lambda_0(\nx,\ny) + \lambda_1(\nx,\ny) \left(\nxy -
              \sqrt{\nx \ny}\right) \ ,
\eeq
which immediately implies
\beq
    \BX = - 2 \left[\frac{\partial \lambda_0}{\partial \nx} +
          \frac{\partial \lambda_1}{\partial \nx} \left(\nxy -
          \sqrt{\nx \ny}\right) - \lambda_1 \frac{n_\y}{2 n_\x}\right]
          \quad , \quad
    \BX_{, \x \y} = - 2 n_\x n_\y \frac{\partial \lambda_1}{\partial \nx} \ ,
\eeq
and
\beq
    \AXY = - \lambda_1 \quad , \quad \AXY_{, \x \y} = 0 \ .
\eeq
We shall make one further simplifying approximation, which is to take
$\lambda_1$ to be constant so that
\beq
    \BX_{, \x \y} = 0 \ .
\eeq
This leaves us with the \underline{single} entrainment parameter $\lambda_1$.

With these approximations we find
\beq
    \left[\begin{array}{cc}
    \BX \left(\sigma^2_\x - c^2_\x\right) & - \Ccc \sqrt{\BX \BY} - \AXY
    \frac{\epsilon_{\x \y} k_\x}{\epsilon_{\y \x} k_\y} \left(\sigma^2_\x - 1
    \right) \\
    - \Ccc \sqrt{\BX \BY} - \AXY \frac{\epsilon_{\y \x} k_\y}
    {\epsilon_{\x \y} k_\x} \left(\sigma^2_\y - 1\right) & \BY \left(
    \sigma^2_\y - c^2_\y\right)
    \end{array}\right]
    \left[\begin{array}{c}
    u^\x_a A^a_\x \\
    u^\y_a A^a_\y
    \end{array}\right] =
    \left[\begin{array}{c} 0 \\ 0 \end{array}\right]
    \ . \label{ccentalg}
\eeq
From \eqref{kmag} and \eqref{kequal} we have
\beq
    \frac{k_\x}{k_\y} = \sqrt{\frac{1 - \sigma^2_\y}{1 - \sigma^2_\x}} \ ,
\eeq
and thus the dispersion relation becomes
\beq
    0 = \left(\sigma^2_\x - c^2_\x\right) \left(\sigma^2_\y - c^2_\y\right)
        - \left[\Ccc + \Cent \sqrt{\left(1 - \sigma^2_\x\right)
        \left(1 - \sigma^2_\y\right)}\right]^2 \ , \label{dispccent2}
\eeq
where
\beq
    \Cent = \epsilon_{\x \y} \epsilon_{\y \x} \frac{\AXY}{\sqrt{\BX \BY}} \ ,
\eeq
and because the inverse of \eqref{momfluxmat} must exist, $|\Cent|
\neq 1$. Note that, since $\sigma^2_\y\rightarrow 1$ as $v_{\x
\y}\rightarrow 1$, $\Cent$ does not affect the ultra-relativistic limit. Hence, 
equation \eqref{ultrarel} [with $\cos(\theta_{\x \y})=\pm 1$] is still
valid in this case.

We stress that, unlike the simpler cases, \eqref{dispccent2} does not
hold for arbitrary propagation direction with respect to the relative
velocity. \ This is important qualitatively and quantitatively, since
it mirrors the fact that entrainment enters the master function in a
fundamentally different way: At first-order in the relative velocity
squared. \ In the dispersion relation, however, entrainment
contributes even in the limit of zero relative velocity, because there
are still two sets of interacting sound waves. \ In fact, we will now
follow the earlier analysis of causality and absolute stability by
taking this limit.

From \eqref{sigtrans} we see $\sigma^2_\x = \sigma^2_\y$ so that
\beq
    \left(\sigma^2_\x - c_\x^2\right) \left(\sigma^2_\x - c^2_\y\right) -
    \left[\Ccc + \Cent \left(1 - \sigma^2_\x\right)\right]^2 = 0 \ .
\eeq
It is particularly instructive to consider the entrainment alone, \ie set
$\Ccc = 0$. \ The corresponding dispersion relation is a quadratic
in $\sigma^2_\x$, and has the solutions
\beq
    \sigma^2_\x = \frac{c^2_\x + c^2_\y - 2 \Cent^2 \pm \left[\left(c^2_\x -
    c^2_\y\right)^2 + 4 \Cent^2 \left(1 - c^2_\x\right) \left(1- c^2_\y\right)
    \right]^{1/2}}{2 \left(1 - \Cent^2\right)} \ . \label{sigxentsln}
\eeq
For $c^2_{\x , \y} \le 1$, the discriminant is obviously positive and hence
the $\sigma^2_\x$ are real. \ In order to analyze absolute stability and
causality we need to consider the ranges $0 \le \Cent^2 < 1$ and $1 <
\Cent^2$ separately. \ We will look at $\Cent^2 > 1$ first.

The first step is to re-write \eqref{sigxentsln} so that the denominator is
positive:
\beq
    \sigma^2_\x = \frac{2 \Cent^2 - c^2_\x - c^2_\y \pm \left[\left(c^2_\x -
    c^2_\y\right)^2 + 4 \Cent^2 \left(1 - c^2_\x\right) \left(1- c^2_\y\right)
    \right]^{1/2}}{2 \left(\Cent^2 - 1\right)} \ . \label{slnge1}
\eeq
Since we are imposing $c^2_{\x \y} \le 1$, and $\Cent^2 > 1$, the terms
outside the square root in the numerator are positive. \ Therefore, the ``+''
solution is absolutely stable. \ But, we can also show that it cannot be
causal. \ As for the ``$-$'' solution, we can easily show that it is
absolutely stable only if $\Cent^2 < 1$, which cannot be satisfied. \ Hence,
$\Cent^2 > 1$ does not lead to both absolute stability and causality, and is
therefore ruled out.

The  range  $0 \le \Cent^2 < 1$ is a different story, because the terms
outside the radical in the numerator of \eqref{sigxentsln} are not of any
definite sign. \ This affects the absolute stability analysis more than the
determination of causality. \ In fact, the causality analysis is sufficiently
straightforward that we will simply state that this requirement is satisfied
for this range of $\Cent^2$. \ In order to assess absolute stability, it is
useful to introduce
\beq
    \tau = \left[\left(c^2_\x - c^2_\y\right)^2 + 4 \Cent^2 \left(1 -
             c^2_\x\right) \left(1 - c^2_\y\right)\right]^{1/2} \ .
\eeq
This allows the numerator of \eqref{sigxentsln} to be rewritten in such a way
that the absolute stability condition becomes
\beq
    \tau^2 - \left[\pm 2 \left(1 - c^2_\x\right) \left(1 - c^2_\y\right)
    \right] + \left[c^2_\x \left(1 - c^2_\y\right) + c^2_\y \left(1 -
    c^2_\x\right)\right] \left[2 - \left(c^2_\x + c^2_\y\right)\right] \le 0
    \ , \label{tauineq}
\eeq
where the ``$\pm$'' corresponds to that of \eqref{sigxentsln}. \ The final
step is to factorize \eqref{tauineq} and thereby obtain
\beq
    \left\{\tau \pm \left[2 - \left(c^2_\x + c^2_\y\right)\right]\right\}
    \left\{\tau \mp \left[c^2_\x \left(1 - c^2_\y\right) + c^2_\y \left(1 -
    c^2_\x\right)\right]\right\} \le 0 \ ,
\eeq
where if the ``$+$'' is taken from the first factor then the ``$-$'' must be
taken in the second, and vice versa. \ In either case, the factor that has
the ``$+$'' is positive definite, and so the other factor must be less than
zero. \ When the first factor takes the ``$-$'' the inequality leads to
$\Cent^2 \le 1$. \ The other choice leads to the more restrictive condition
of $\Cent^2 \le c^2_\x c^2_\y$.

To summarize, we have shown that when $\Cent^2 > 1$, there is either no
absolute stability or causality, which makes this range unphysical. \
Meanwhile, for $0 \le \Cent^2 \le c^2_\x c^2_\y$ the system is causal and
absolutely stable. \ In terms of our earlier definitions, this translates into
\beq
    ({\AXY})^2\ \le \BX \BY \left(\frac{\partial \log \BX}{\partial
    \log n_\x} + 1\right) \left(\frac{\partial \log \BY}{\partial \log n_\y}
    + 1\right) \label{entconst}
\eeq
as a constraint on the master function, when the relative speed between the
two fluids is sufficiently small.

\begin{figure}
  \includegraphics[height=0.6\textwidth,width=0.9\textwidth,clip]
   {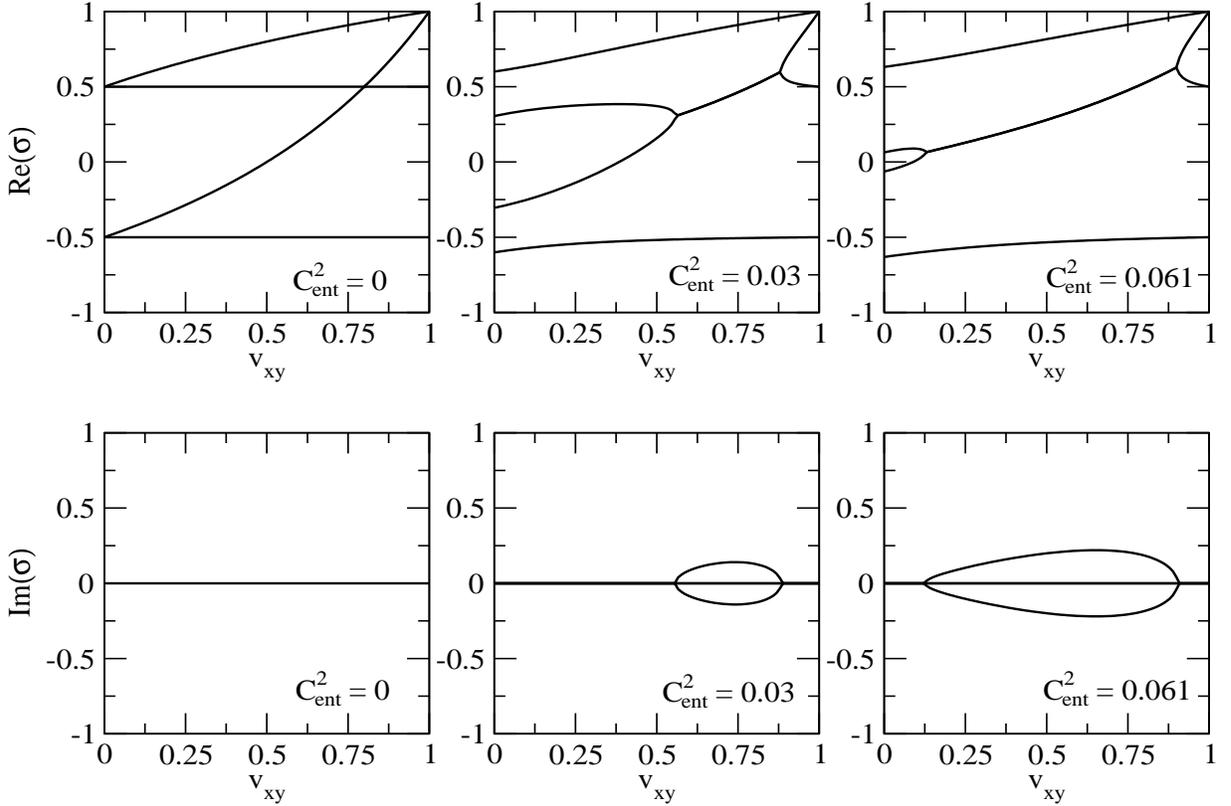}
   \caption{Plots of the real and imaginary parts of the mode frequencies
   $\sigma_\x$ as functions of $y$ and $\Cent^2$, for $c_y = 0.5$, and
   $b^2 = 1.0$. \ The lines of intersection and subsequent merger of two
   frequencies signal the presence of a two-stream instability.}
   \label{entplt}
\end{figure}

Finally, we turn to a numerical/graphical analysis for exposing the two-stream
instability due to entrainment coupling. \ As in the cross-constituent
coupling case, we use the re-scalings of \eqref{rescale}, except that
$\Cent^2$ replaces $\Ccc^2$ in $a^2$. \ Equation \eqref{dispccent2} becomes
\beq
   \frac{x^2 - b^2}{a^2} \left[\frac{\left(x - y\right)^2}{\left(c^2_\y y x -
   1\right)^2} - 1\right] = \left(c^2_\y x^2 - 1\right) \left[\frac{c^2_\y
   \left(x - y\right)^2}{\left(c^2_\y y x - 1\right)^2} - 1\right] \ .
   \label{centquartic}
\eeq
The parameter values are restricted to those that maintain absolute stability
and causality. \ Fig.~\ref{entplt} plots the real and imaginary parts of the
four solutions to \eqref{centquartic}. \ As before, the solutions for
$\sigma_\x$ are taken to be functions of the relative flow parameter $y$ and
the coupling $\Cent^2$, with $c_\y = 0.5$ and $b^2 = 1.0$. \ Fig.~\ref{entplt}
is not so dissimilar from what we find for cross-constituent coupling, thus
highlighting that the instability is not sensitive to the type of coupling
between the two fluids.

\section{Closing Remarks}
\label{conclu}

There are several examples of relativistic systems that require multi-fluid
dynamics for qualitative understanding and quantitative accuracy. \ In such
systems different interpenetrating fluid components (eg.~particles and
entropy in heat conducting situations or a superfluid condensate and finite
temperature excitations) can flow with distinct velocities.

We have examined plane-wave propagation for a generic two-fluid system. \ By
imposing the constraints of absolute stability (necessary in order for the
components not to separate already in the absence of flow) and causality we
have established limits on the equation of state (as represented by the
master function $\Lambda$). \ In particular, we place constraints on the free
sound speeds, the cross-constituent coupling and the entrainment. \ Some of
the obtained results are more or less trivial extensions of the single fluid
result, but others are unique to the two-fluid problem. \ The condition
\eqref{entconst} on the entrainment is a new result, and as such serves as a
new condition on, say, the kind of $\sigma - \omega$ model used by Comer and
Joynt \cite{comer03:_rel_ent} to model entrainment in the outer cores of
neutron stars.

We have demonstrated (for the first time) the existence of a relativistic
two-stream instability. \ This is a generic phenomenon, that does not require
particular fine-tuning to be triggered, nor is it limited to any specific
physical system. \ The only requirement is that there is a relative
(background) flow and some type of coupling between the fluids. \ While it is
true that a single fluid can have an analogous instability, it is only active
at an interface where there is shearing motion. \ Our analysis assumes that
the two fluids are interpenetrating.

In order to exhibit the generic nature of the two-stream instability, we have
kept the analysis rather abstract. \ On the one hand, this means that it
should be relatively straightforward to apply our results to particular
physical systems. \ On the other hand, it means that we have not yet
discussed the relevance of the instability for any particular system. \ A
more detailed consideration of multi-fluid problems in relativity is required
in order to establish whether this mechanism operates in nature. \ There are
already exciting results that hint at this class of instabilities being
associated with pulsar glitches
\cite{andersson04:_twostream,glampedakis09:_glitch_prl}. \ In addition to
exploring possible situations where these instabilities may operate, it would
be very interesting to probe the nonlinear development of the unstable waves.
\ So far, all studies have been at the linear perturbation level. \
The results establish the presence of the instability, but they do not shed
any light on what happens once the unstable oscillation reaches nonlinear
amplitudes. \ Detailed studies of this problem are essential if we
are to understand the actual dynamical role of this instability.

\acknowledgments

NA acknowledges support from STFC via grant number PP/E001025/1.
CLM is supported by CONACyT. GLC acknowledges partial support from NSF via 
grant number PHYS-0855558.


\end{document}